# Remarks on the treatments of non-solvable potentials

**Bülent Gönül and Yücel Cançelik**

University of Gaziantep, Department of Engineering Physics, 27310 Gaziantep-Turkey

The recently introduced scheme [20,21] is extended to propose an algebraic non-perturbative approach for the analytical treatment of Schrödinger equations with non-solvable potentials involving an exactly solvable potential form together with an additional piece. As an illustration the procedure is successfully applied to the Cornell potential by means of very simple algebraic manipulations. However, instead of providing numerical eigenvalues for the only consideration of the small strength of the related linear potential as in the previous reports, the present model puts forward a clean route to interpret related experimental or precise numerical results involving wide range of the linear potential strengths. We hope this new technique will shed some light on the questions concerning with the limitations of the traditional perturbation techniques.

## I. INTRODUCTION

The Schrödinger equation with the Cornell potential, $V(r) = -\frac{a}{r} + br$, also known as the Coulomb plus linear potential, has received a great deal of attention [1-19] as an important non-relativistic model in both particle physics, more precisely in the context of meson spectroscopy where it is used to describe systems of quark and antiquark bound states, and in atomic and molecular physics where it represents a radial Stark effect in hydrogen. Aside of the physical relevance, the solutions of the Schrödinger equation for the Coulomb plus linear potential have been rigorously investigated with a large number of techniques due to its also nontrivial mathematical properties. In addition, this potential has an advantage that leads naturally to two choices of parent Hamiltonian in perturbative treatments, one based on the Coulomb part and the other on the linear term, which can be usefully compared. Although such models have been studied for many years, exact solutions of Schrödinger equation with this potential are still unknown to a great extent. Most of the earlier work either relies on direct numerical integration of the Schrödinger equation or various techniques for approximating the eigenenergies.

Within this context, and also gaining confidence from the succesful applications [20,21] of the recently developed formalism appeared in a series of papers for the analytical treatment of exactly- and quasi-/conditionaly-exactly solvable potentials used in analysing heavy quarkonium observations, we aim in this letter at indicating how the previous recipe in [20,21] can be extended to give a further novel prescription for also non-exactly solvable more realistic potentials employed in different disciplines of physics and engineering as well. This work therefore is the completeness of the preceding reports in [20,21] and the justification of the flexibility of the model developed and used for various treatments. The method however we propose in this article do not suggest a new prescription for the energy eigenvalues of such potentials, unlike the previous works. Approaching to such problems in different perspective, we will try here, in contrary to the previous reports, to suggest an unusual but simple expression for an explicit analysis of the results obtained exactly, but without providing a deep physical interpretation of the findings, such as precise numerical treatments or reliable experimental observations for any physical problem that may be modelled by this class of potential. More specifically, the present communication emphasizes an interesting relationship between the critical $r$ – values, leading to exact energies, in the problem considered and the limitations of the known perturbation theories which depend on finding an appropriate parent Hamiltonian for which no general procedure is available even though the choice may be crucial to the success of such approximate schemes. In this respect, the comments on our tabulated results provide intuition on this crucial choice, which seems to have been insufficiently appreciated hitherto.

The paper is organized as follows. In section 2, we extend the scenario in [20,21] to $N$-dimensional space for the treatment of non-solvable potentials. The main features of the formalism used and its application to the Cornell type potentials are given in section 3 and finally, in section 4 we draw our conclusions.

## 2. FORMALISM

Here, for the stringent test of the present model with the others available in the literature, the usual one-dimensional solution of non-relativistic Schrödinger equation is extended to arbitrary dimensions considering the frame in [20,21].

The radial Schrödinger equation for a spherically symmetric potential in $N$-dimensional space $(\hbar = 1)$ reads [22]

$$\frac{d^2 R}{dr^2} + \frac{N-1}{r}\frac{dR}{dr} = 2m\left[\left(V(r) + \frac{\ell(\ell+1)}{2mr^2}\right) - E\right]R \ , \tag{1}$$

which is transform to

$$\frac{d^2\Psi}{dr^2} = 2m\left[\left(V(r) + \frac{(M-1)(M-3)}{8mr^2}\right) - E\right]\Psi \ , \tag{2}$$

where $\Psi(r) = r^{(N-1)/2} R(r)$, being the reduced radial wavefunction, and $M = N + 2\ell$. Eq. (2) can also be expressed as

$$\frac{d^2\Psi}{dr^2} = 2m\left[\left(V(r) + \frac{(\Lambda)(\Lambda+1)}{2mr^2}\right) - E\right]\Psi \ , \tag{3}$$

where $\Lambda = (M-3)/2$. We see that the radial Schrödinger equation in *N*-dimensions has the same form as the three-dimensional one. Consequently, given that the potential has the same form in any dimension, the solution in three dimensions can be employed to obtain the solution in any dimension simply by using the substitution $\ell \to \Lambda$.

At this stage, we extend our previous formalism assuming, in Eq. (3), that $\Psi(r) = F[g(r)] f(r)$ where $F(g)$ yields an algebraic closed solution for an exactly solvable potential with $F(g)$ being a special function satisfying a second-order differential equation

$$\frac{d^2 F}{dg^2} + Q(g)\frac{dF}{dg} + R(g) F(g) = 0 \ , \tag{4}$$

while $f(r)$ is the moderating function in connection with a perturbing/modifying piece of the entire potential given. The form of $Q(g)$ and $R(g)$ is already well defined [23] for any special function $F(g)$ when dealing with analytically solvable potentials. However, in case of a realistic non-solvable problem consideration one should derive a reliable scheme to obtain the corrections more accurately due to the moderating piece of the potential in the ligh of exact part of the calculations. This is the significant point in the framework of the new formalism to reach physically meaningful solutions.

After some algebra, Eq. 3 turns out to be

$$\frac{F''}{F}(g')^2 + 2\frac{F'}{F}\frac{f'}{f}g' + \frac{F'}{F}g'' + \frac{f''}{f} = 2m(V_{eff} - E) ,  \qquad (5)$$

where $V_{eff}(r) = V(r) + \frac{\Lambda(\Lambda+1)}{2mr^2}$ and the primes denote derivatives with respect to $r$, except the ones related to $F$ as $F' = \partial F/\partial g$. To make a relation between Eqs. (4) and (5), we transform the above equation to the form of

$$F'' + F'\left(\frac{2}{g'}\frac{f'}{f} + \frac{g''}{(g')^2}\right) + F\left[\frac{f''}{f(g')^2} + \frac{2m}{(g')^2}(E - V_{eff})\right] = 0 . \qquad (6)$$

Bearing in mind that the full spectra of the system underlined is given by $E = E_{ES} + \Delta E$, Eq. (6) can be transformed easily to a coupled equation [20,21] leading to the explicit solutions for exactly solvable part of the potential

$$E_{ES} - \left[V_{ES}(r) + \frac{\Lambda(\Lambda+1)}{2mr^2}\right] = \frac{(g')^2}{2m}R(g) \qquad (7)$$

and

$$2\left(\frac{F'g'}{F}\right)\frac{f'}{f} + \frac{f''}{f} = 2m[\Delta V(r) - \Delta E] \qquad (8)$$

which is responsible for the computation of the rectifications. More clearly, the solution of (8) in either an explicit or approximate form depending on the structure of $\Delta V$, provides us to see the modifications $(\Delta E, f)$ to the exact solutions $(E_{ES}, F(g))$ due to the consideration of a modifying/perturbing interaction $(\Delta V)$ in a realistic application.

## 3. APPLICATION

As the main aim of this article is in general to obtain a physically reliable scheme for more accurate solutions of non-solvable potentials, and also to illustrate the flexiable applicability of the model, we focus here a specifally chosen example that is the Cornell potential in arbitrary dimensions,

$$V_{eff}(r) = V_{ES}(r) + \Delta V(r) = \left(-\frac{a}{r} + \frac{\Lambda(\Lambda+1)}{2mr^2}\right) + br \ , \qquad (9)$$

in which the first term is an exactly solvable Colulomb-like potential that has the well known solutions [21] along the frame of (7) which provides a safe basis for the computation of the connected solutions

$$E_n^{ES} = -\frac{ma^2}{2(n+\Lambda+1)^2} \quad , \quad \Psi_n^{ES}(r) \sim F_n(g) = \left[e^{-g/2} g^{(\alpha+1)/2} L_n^\alpha(g)\right] \qquad (10)$$

for which we have an appropriate choice [23] such as

$$Q(g) = 0 \quad , \quad R(g) = \frac{n+\Lambda+1}{g} - \frac{\Lambda(\Lambda+1)}{g^2} - \frac{1}{4} \ , \quad n = 0,1,2,..., \qquad (11)$$

where $g(r) = \frac{2mar}{n+\Lambda+1}$, $\alpha = 2\Lambda+1$, and finally $L_n^\alpha(g)$ is the generalized Laguerre polynomial. For the relation between $f(r)$, $g(r)$ and $Q(g)$, the reader is referred to [20, 21].

Obviously, the task here is to solve Eq. (8). Tracking down approximate forms of its solutions has always aroused interest [1-19]. Apart from being useful in understanding of many physical phenomena, the importance of searching for them also stems from the fact that they very often provide a good starting point for undertaking perturbative calculations of more complex systems. Considering all these earlier works and the experience gained from such calculations, we propose below in section 3.2 an alternative simple receipe, having a physically motivated visualizable framework, which enables one to give secure physical explanations behind the precise numerical findings or experimental measurements concerning with the interaction of interest.

## 3.1. Attempts for the solution of Eq. (8)

In the last three decades with the rapid development of nonlinear science, there has appeared ever increasing interest of scientists and engineers in the analytical techniques for nonlinear problems. In particular, most of the one-particle potentials that are encountered in such quantum mechanical applications do not allow closed exact solutions either for the energy eigenvalues or for the wavefunctions as in the present case. In these circumstances, one resorts to approximate methods which may be suitable to the particular situation to obtain approximate solutions. Perhaps the most useful solutions in this context are the perturbative solutions. With this consideration, we first show that Eq. (8) may also be transformed to a perturbative scheme. In this case, reminding the full wavefunction description $\Psi = F(g)f$ used through the article, one needs to rearrange Eq. (8) as $F_n(g)f_n'' + 2(F_n'g')f_n' = 2m[\Delta V(r) - \Delta E]F_n(g)f_n$. Using the spirit of perturbation theories, $f(r)$ and $\Delta E$ are expanded in terms of the perturbation parameter $\lambda$ such as $f_n(r,\lambda) = 1 + \sum_{j=1}^{\infty} \lambda^j f_{nj}(r)$, $\Delta E_n(\lambda) = \sum_{j=1}^{\infty} \lambda^j \Delta E_{nj}$, $\Delta V(r) = \lambda^1(br)$ where $j$ denotes the perturbation order. Substitution of these expansions into the new form of Eq. (8) above and by equating terms with the same power of $\lambda$ on both sides yield, interestingly, the old but well-known alternative perturbation procedure in [24]. However our exhaustive application results, which are not given here for the sake of clarity, have revealed once more that such approaches inherently work well for only quite small parameters $(b \ll 1)$ of the linear potential in (9).

We have then drawn our attention to the supersymmetric perturbation theory (SSPT) [25], a lesser-known alternative to the usual perturbative treatments in quantum mechanics. Therefore, in addition to our first attempt mentioned above, the whole formalism discussed in this article has then been expressed within the framework of SSPT defining the forms of the superpotentials in there as $W_n^{ES} = -\frac{1}{\sqrt{2m}}\frac{F_n'g'}{F_n(g)}$, corresponding to the exact (Coulomb like) potential, and $\Delta W_n = -\frac{1}{\sqrt{2m}}\frac{f_n'}{f_n}$ concerning the modifying (linear potential) solutions, for the alternative treatment of Eq.(8). Later, as the next step, the familiar expression of the supersymmetric quantum mechanics $(W_n^{ES} + \Delta W_n)^2 - (W_n^{ES} + \Delta W_n)'/\sqrt{2m} = V - E_n$ has been splitted in two parts as $\left(W_n^{ES}\right)^2 - \left(W_n^{ES}\right)'/\sqrt{2m} = V_{ES} - E_n^{ES}$ and $2W_n^{ES}\Delta W_n + \Delta W_n^2 - \left(\Delta W_n\right)'/\sqrt{2m} = \Delta V - \Delta E_n$ generating strictly Eqs. (7) and (8), respectively, for individual quantum states. Nevertheless, the expansion of $\Delta W$ as described in [25], together with the proper expansions of $\Delta V$ and $\Delta E$ at each succesive perturbation order for the approximate solution of the refinements makes the calculation scheme considerably cumbersome due to a tedious iterative procedure used for the description of the partner potentials in the model. Beside of this inconvenience, which is not illustrated here as it is out of the scope of the present work, we have seen again that this type of calculations work only rather small confining potential parameter. These two points make the calculation scheme naturally useless and impracticable as in our previous attempt.

To sum up, with the experience gained from these our earlier calculations and also taking full advantages of the approaches in Refs. [24,25] we present here a more economical but instructive prescription for the interpretation of accurate solutions of Eq. (8) in connection with the analysis of Cornell like funnel potentials.

### 3.2. Alternative approach to Eq. (8)

Clearly, the total wavefunction of the system under consideration assumes its asymptotic behaviour at very large distance that is completely determined by the modifying linear piece $(br)$ in the full potential appeared in (9), which is the same for all (ground and excited) states, while at intermediate distances the wavefunction still decays exponentially but now governed by the Coulomb term. Hence, the lack of a unique frame for the equal treatments of the main and modifying parts of the interaction potential considered in perturbation theories, unlike the model presented here, does not decisevly affect the description of especially the ground-state with small parameters, but destroys the ability of the potential to describe such interactions beyond small perturbation domains. In this respect, the present consideration would offer the opportunity to improve such calculations due to the clear visualization of the refinements, in an explicit manner, in connection with the confinement potential term.

As the full solution for the Coulomb like potentials at intermediate distances of the present interaction is known, Eq. (10), and the both sides of Eq. (7) are equal to each other in the asymptotic region due to the structure of $R(g)$ in (11) when $r \to \infty$, clarifying the non-contribution of the Coulomb potential at large distances, one needs to focus at just Eq. (8) to obtain the corrections appeared in the asymptotic domain. In this case, evidently the unnormalized solution in this region is given by the Airy function as in Ref. [19], $f(r) = Ai\left((2mb)^{1/3} r\right)$, if (considering only the ground-state for the present purpose)

$$\Delta E_{n=0} \simeq -\frac{1}{m}\left(F'_{n=0}\, g'/F_n\right) f'/f \quad \Rightarrow \quad f''(r) - (2mbr) f'(r) = 0 \tag{12}$$

Thus, this assumption guides us to define the entire solution as $\Psi_{n=0}(r) = F_{n=0}(g) f(r) = \Psi^{ES}_{n=0}(r) Ai\left((2mb)^{1/3} r\right)$ where $\Psi^{ES}_{n=0}$ is given by (10).

In this circumstance, the correction term for the energy contribution in arbitrary dimensions takes the following form

$$\Delta E_{n=0} = \left(\frac{a}{\Lambda+1} - \frac{\Lambda+1}{m\, r}\right) \frac{f'(r)}{f(r)} \tag{13}$$

which depends on $r$. This peculiar behaviour is of course due to the present consideration of the non-exactly solvable nature of the Cornell potential. Before proceeding, however, it is worthwhile to pay some attention on (13). Recalling our previous second attempt to propose an approximate scheme for the solution of (8) within the frame of the supersymmetric perturbation theory, that has been discussed in the previous section, it is clear that

$$2W^{ES}_n \Delta W + \left(\Delta W_n^2 - \frac{(\Delta W)'}{\sqrt{2m}}\right) = br - \Delta E_n \tag{14}$$

where $W^{ES}_n = -F'_n g'/\sqrt{2m} F_n(g)$ and $\Delta W = -f'/\sqrt{2m} f$ for the present treatment case. The substitution of these superpotentials into (14) yields explicitly Eq.(8) and subsequently Eqs. (12) and (13) for $n=0$ consideration, as $\Delta V = \Delta W^2 - (\Delta W)'/\sqrt{2m} = f''/2mf = br$ and

$$\Delta E_{n=0} = 2W^{ES}_{n=0} \Delta W = \left(\frac{a}{\Lambda+1} - \frac{\ell+1}{m\, r}\right) \frac{f'(r)}{f(r)} \tag{15}$$

From (15), one sees that a proper choice of $\Delta W$ in connection with the pertinent wavefunction $(f)$ yields a constant value for $\Delta E$ reducing the problem to an analytically solvable case. Otherwise, as in the present non-solvable potential consideration, $\Delta W$ and $\Delta E$ should be expanded as in either the usual perturbation or supersymmetric perturbation theories to solve the problem in an approximate form. Unfortunately, such treatments have some drawbacks mentioned in 3.1.

Therefore, instead of removing this apparent disadvantage in (15) using an iterative or perturbative scheme, we adopt here a different strategy concerning with the clarification of the physics behind nearly precise numerical calculations in the light of Eq. (13). This will reveal an interesting relation between the critical $r$-values in the wavefunction, yielding exact energies, and the choice of an appropriate parent potential in perturbation like theories. A quick analysis of (13), at this stage, tells us that if $r = r_0 = \frac{(\Lambda+1)^2}{m\,a}$ then $\Delta E = 0$, in which condition all the refinements disappear, reducing the entire system to the purely exactly solvable Coulombic case where $E \to E_{n=0}^{ES} = -a/2r_0 = -ma^2/2(\Lambda+1)^2$. This location in the corresponding energy expression is in fact the maximum of the ground-state Coulomb wavefunction, $\Psi_{n=0}^{ES} = r^{\Lambda+1} \exp\left(-\frac{ma}{\Lambda+1}r\right)$, that can be easily checked through $\left(\partial \Psi_{n=0}^{ES}/\partial r\right) = 0$.

From this short instructive discussion, a good choice is to proceed with the another maximum $(r_0)$ in connection with the the ground-state of the full wavefunction in the asymptotic domain, $\Psi \sim r^{\Lambda+1} f(r)$, for the precise calculations of $\Delta E$. In fact, for the Cornell potential, because of the attractive Coulomb term, the potential function is in general not bounded below and therefore the choice of $r_0$ to be the location of the maximum of the ground-state wavefunction in the asymptotic region, as a initial guess, seems reasonable. Hence, one should expect the physically resonable $r$-values in (13), responsible for the exact energies, in the vicinity of this maximum point. To define exactly the positions of these shifted $r = r_{\Delta E}$ values, we need to consider the exact energies calculated through precise numerical techniques such as the ones in Refs. [13,19] and the related references therein. As $\Delta E = E_{exact} - E_{ES}$, Eq. (13) can be expressed as

$$E_{exact}^{n=0} = -\frac{ma}{2(\Lambda+1)^2} + \left(\frac{a}{\Lambda+1} - \frac{\Lambda+1}{mr_{\Delta E}}\right)\frac{f'(r)}{f(r)} \tag{16}$$

which serves us to find exact locations for $r-$ values related to the exact energy eigenvalues of the individual $\ell-$ states. The systematic calculation of $r_{\Delta E}$ value via (16) offers no difficulty if we resort a computer algebra system like Mathematica, Mapple or Reduce.

The results obtained are tabulated through the three tables. For the purpose of consistency, we have calculated each $r_{\Delta E}$ value to 16 significant figures, unlike the other two values rounded in there for clarity. In Tables 1 and 2 we report the eigenvalues for the Schrödinger equation in distinct dimensions with the potential $V = -1/r + br$ considering the different sizes of the linear potential strength. Table 3 illustrates $S-$ wave ground-state eigenenergies for the potential $V = -a/r + r$ with dissimilar values of the parameter $a$. From these results, the following comments apply:

1- All $r_{\Delta E}$ − values obtained, in $1/GeV$ units, satisfy $r_{\Delta E} \prec \dfrac{(\Lambda+1)^2}{m\,a}$ condition, as $\Delta E$ is always positive whereas $f'/f$ has a negative structure in Eq. (13), except $a = 0$ case.

2- The first $r_{\Delta E}$ − values in Tables 1 and 2 for $b = 0.01$ reveal why one naturally expects a bound only for this situations, which are very close to the critical $r-$ value, $r = r_0 = (\Lambda+1)^2/ma$, where the Coulomb like potential is dominant and the linear part of the Cornell potential may be treated as a perturbation in this case.

3- From Table 1, as $r_0$ and $r_{\Delta E}$ depend on $b-$ and $\ell-$ values due to the form of the related function, $r^{\ell+1} f(b,r)$, in the asymptotic region, the decrease in these $r-$ values can be readily observed through the increase in the parameter $b$. However, a considerable increment in $r_0$ and $r_{\Delta E}$ are clearly visible when $\ell$ increases for each individual $b-$ value. Since the asymptotic solution is independent of the Coulombic parameter $a$, the $r-$ values of interest are not affected by the changes in $a$ as in Table 3.

**Table [1]:** Ground-state $r_{\Delta E}$ − values for $V(r) = \dfrac{-1}{r} + br$ in $N=3$ dimensional space.

| $b$ | $\ell$ | $r_0$ | $r_{\Delta E}$ (Eq.13) | $\Delta E_{n=0}$ |
|---|---|---|---|---|
| 0.01 | 0 | 4.103 | 1.7436481087936350 | 0.029 |
|  | 1 | 6.873 | 4.9460678940048420 | 0.080 |
|  | 2 | 9.195 | 7.9252834202633755 | 0.130 |
|  | 3 | 11.265 | 10.555231364572375 | 0.175 |
|  | 4 | 13.163 | 12.921702562396474 | 0.216 |
|  | 5 | 14.935 | 15.094068452133266 | 0.254 |
| 1 | 0 | 0.884 | 0.7994448794104599 | 1.648 |
|  | 1 | 1.481 | 1.5319979780777233 | 2.888 |
|  | 2 | 1.981 | 2.1271924940550924 | 3.878 |
|  | 3 | 2.427 | 2.6476825358214430 | 4.742 |
|  | 4 | 2.836 | 3.120002415656866 | 5.527 |
|  | 5 | 3.218 | 3.5579114353323114 | 6.255 |
| 100 | 0 | 0.190 | 0.20718831032409812 | 46.652 |
|  | 1 | 0.319 | 0.3568814759026067 | 70.079 |
|  | 2 | 0.427 | 0.48025853095063736 | 89.743 |
|  | 3 | 0.523 | 0.5892382922692437 | 107.350 |
|  | 4 | 0.611 | 0.6887636263960271 | 123.572 |
|  | 5 | 0.693 | 0.7814337345649496 | 138.768 |

**Table [2]:** Ground-state $r_{\Delta E}$ – values for $V(r) = \dfrac{-1}{r} + br$ in $N = 4$ dimensional space.

| $b$ | $\ell$ | $r_0$ | $r_{\Delta E}$ | $\Delta E_{n=0}$ |
|---|---|---|---|---|
| 0.01 | 0 | 5.566 | 3.3312700304565390 | 0.534 |
| | 1 | 8.074 | 6.4834209674432110 | 0.106 |
| | 2 | 10.255 | 9.278604516903260 | 0.153 |
| | 3 | 12.232 | 11.766427983937469 | 0.196 |
| | 4 | 14.063 | 14.028816948623492 | 0.235 |
| | 5 | 15.783 | 14.585313802191063 | 0.293 |
| 1 | 0 | 1.199 | 1.1896870585180375 | 2.314 |
| | 1 | 1.739 | 1.8415039963613402 | 3.404 |
| | 2 | 2.209 | 2.394675709114608 | 4.322 |
| | 3 | 2.635 | 2.888823024670750 | 5.143 |
| | 4 | 3.030 | 2.4699633437009440 | 7.094 |
| | 5 | 3.400 | 2.3228580785446677 | 8.805 |

4- When $b$ increases $\Delta E$ gets larger values. In particular in case of $b = 100$ in Table 1, where the linear confinement is more active, the energy corrections are striking with the rise of $\ell$ – values. In this circumstance, the Colombic piece may be regarded as a perturbative term. In contrast to this observation, Table 3 shows that $\Delta E$ decreases with increasing $a$ – values which makes the Coulomb potential more stronger than the confining part. Consequently, the refinements in eigenvalues due to the linear potential diminish as expected. These results confirm the comments given above.

**Table [3]:** The lowest level $r_{\Delta E}$ – values for $V(r) = \dfrac{-a}{r} + r$ in $N=3$ dimension with $r_0 = 0.884$

| $a$ | $r_{\Delta E}$ | $\Delta E_{n=0}$ | $a$ | $r_{\Delta E}$ | $\Delta E_{n=0}$ |
|---|---|---|---|---|---|
| 0.0 | 1.0092498710582083 | 2.338 | 1.0 | 0.7994448794104599 | 1.648 |
| 0.1 | 0.9871174720215355 | 2.254 | 1.1 | 0.7806100091335577 | 1.593 |
| 0.2 | 0.9650736643159619 | 2.177 | 1.2 | 0.7622476487440810 | 1.541 |
| 0.3 | 0.9432041027784062 | 2.101 | 1.3 | 0.7443630403947710 | 1.491 |
| 0.4 | 0.9215784931711197 | 2.028 | 1.4 | 0.7269585604248531 | 1.443 |
| 0.5 | 0.9002533954125955 | 1.958 | 1.5 | 0.7100341340504672 | 1.397 |
| 0.6 | 0.8792744720957341 | 1.891 | 1.6 | 0.6935875917754227 | 1.353 |
| 0.7 | 0.8586783015732216 | 1.826 | 1.7 | 0.6776149777440374 | 1.311 |
| 0.8 | 0.8384938486996074 | 1.764 | 1.8 | 0.6621108181210410 | 1.270 |
| 0.9 | 0.8187436656830261 | 1.705 | 1.9 | 0.6470684263448296 | 1.232 |

5- Additionally, there is an impressive correlation between the $r$ – values underlined. In Table 1, $r_0 \succ r_{\Delta E}$ is valid for $b = 0.01$ through $\ell = 0$ to $\ell = 4$. The rapid rise in $r_{\Delta E}$ due to increase in $\ell$ – values makes them comparable, even larger than $r_0$ as in the case of $\ell = 5$. This can be understandable as the angular momentum barrier with large $\ell$ provides more positive contribution to the whole potential, which decreases effects of the Coulombic part while the modifications in $\Delta E$ gets larger. On the other hand, for larger values of $b$, $r_{\Delta E} \succ r_0$ through the all $\ell$ – values except only $b = 1$ with $\ell = 0$ case where both potentials are comparable. From this short discussion, we conclude that $r_0 \succ r_{\Delta E}$ signalizes the domains where the Coulombic part is dominant while the linear piece behaves as a perturbating potential whereas $r_{\Delta E} \succ r_0$ indicates the regions where the confining potential is more active and the Coulomb like potential can be treated as a modifying term. Roughly speaking, $r_{\Delta E}$ measures the distance at which the potential changes from being dominantly Coulombic $(r_{\Delta E} \prec r_0)$ to dominantly linear $(r_{\Delta E} \succ r_0)$. In sum, as in a perturbation procedure the value of the model parameters play a crucial role in choosing the parent and perturbative terms, reliable estimations of $r_{\Delta E}$ would be helpful in providing perception on this choice and on the energy corrections as well through Eq. (13). Some such physical insight is usually necessary in any approach which is based on a judicious choice of the either related potential functions or concerning wavefunctions. Within this context, the present comments on our findings would be helpful on clarifying the questions concerning with the limitations of the traditional perturbation techniques.

6- The situation is nonetheless not similar as the dimension increases. In particular, considering $N=4$ case for $b=1$ with $\ell=4$ and $\ell=5$ in Table 2, the trend in $r$-values is opposite to those of $N=3$. The change in the definition of the angular momentum concept in this dimension where $\ell \to \ell+1/2 = \Lambda$, which has been debated in section 2, causes this conflicting behaviour due to the new shifted larger shapes of the Coulomb wavefunction, $r^{\ell+3/2}e^{-mar/(\ell+3/2)}$, and of the concerning asymptotic wavefunction, $r^{\ell+3/2}f(r)$, for the ground-state solutions in this larger dimension. In addition, unlike the case in Table 1, Columbic interaction seems more powerful than the linear end of the potential for even $\ell=5$ in case of $b=1$ shown in Table 2, as still $r_0 \succ r_{\Delta E}$ because of the same reason regarding the another forms of the wavefunctions.

In sum, the comments above clarify why the usual perturbation treatments, such as the ones discussed in section 3.1, do not work for the large $b-$ and $\ell-$values through the analysis of the Cornell potential, in which the Coulomb potential is admitted as always dominant to the linear part of the entire potential. Moreover, Aitchison and Dudek in Ref.[26] put an argument about the significance of some critical $r$-values in the analysis of heavy quarkonium spectra and showed that with the Coulombic part, as parent in the Cornell potential, a bottomonium spectra are well explained than charmonium whereas charmonium states are well elucidated with linear part as the main piece of the same potential, which makes the the whole analysis above noteworthy in this context.

We finally note that the discussion of excited states within the frame of the present scenario is unnecessary as the main points of the work presented through this article, which has been clarified through the tables and comments above, can be safely given only by the consideration of the ground state calculations. However for the treatments of $n \succ 0$ levels within the framework of a similar procedure, the reader is refered to [21].

## 4. CONCLUDING REMARKS

We have presented an alternative technique for the solution of the Schrödinger equation with non-solvable potentials having an analytically soluable part, together with a modifying piece. Although we have limited ourselves to one illustrative example, the range of application of the method is rather large and appears to be straightforward. The formalism has a structure for the analysis of related experimental results and of precise numerical solutions, which can be accomplished with ease. Through the formalism used we have shown how estimates of $r_{\Delta E}$ could help to provide intuition on whether a given state is likely to be modelled better by the Coulomb or the linear part of the Cornell potential, together with reasonable predictions for the energy refinements with the help of Eq. (13), which are significant in the analysis of heavy quarkonium spectra. In view of the importance in analysing such corrections in a simple manner, we believe that the present model would serve as a useful toolbox to interpret even properly chosen more realistic situations, which now occur in experimental observations with the advent of the quantum technology. Additionally, when the structure of a critically stable quantum system is analyzed, understanding the analytic behavior of the solution as a function of different physical parameters is often of decisive importance. This fact justifies the significance of the analytical treatments, such as the one discussed here, in different disciplines of the science.

Aside of the problem posed and discussed through the present article having its own inner mathematical beauty and ability for providing a good starting point for doing calculations perturbatively/non-perturbatively for more complex systems, the present work once more reveals that a proper perturbation method, requiring no small parameters in the equations, which can readily eliminate the limitations of the traditional perturbation techniques which often fails miserably for systems on the border of stability, is urgently required as most nonlinear equations have no small parameter at all. Within this context, nonlinear analysis methods involving homotopy perturbation techniques [27-29] and the quasilinearization method [30-35] which approximates solution of nonlinear differential equations by treating the nonlinear terms as a perturbation about the linear ones, are promising. These models are iterative but not perturbative and give generally stable solutions to nonlinear problems without depending on the existence of a smallness parameter. The use of such powerful treatments for the solution of Eq. [8] within the present framework and its extension to other similar potentials is underway.